\documentclass[12pt]{article}

\usepackage{fancyhdr}
\usepackage{epsfig}
\usepackage{cite}
\usepackage{graphicx}
\usepackage[centertags]{amsmath}
\usepackage{theorem}

\usepackage{youngtab}
\usepackage{graphicx}
\usepackage{amssymb}
\usepackage{latexsym}
\usepackage{epic}
\usepackage{pstricks}
\usepackage{color}
\usepackage{mathrsfs}
\usepackage{latexsym}
\usepackage{cancel}
\usepackage{epic}
\usepackage{braket}
\usepackage{slashed}
\newcommand{\be}{\begin{equation}}
\newcommand{\ee}{\end{equation}}
\newcommand{\bea}{\begin{eqnarray}}

\newcommand{\eea}{\end{eqnarray}}

\numberwithin{equation}{section}
\def\hybrid{\topmargin 22pt    \oddsidemargin 0pt %%%%%%%%%%%%%% Archive-30pt
      \headheight 0pt \headsep 0pt
      \textwidth 6.5in        
       \textheight 9in         
      \marginparwidth .875in
      \parskip 5pt plus 1pt   \jot = 1.5ex}

\hybrid
\usepackage{caption} 
\newcommand{\refe}[1]{Eqn.~(\ref{#1})}

\renewcommand{\thefootnote}{\fnsymbol{footnote}}
\begin{document}
\center{
\begin{flushright} 
DESY 12-128
\end{flushright}
\vspace{1cm}
\begin{center}
{\Large\textbf{General Resonance Mediation}}
\vspace{1cm}

\textbf{Moritz McGarrie$^{1,}$\footnote[2]{\texttt{moritz.mcgarrie@desy.de}}
% and Gianni Tallarita$^{2,}$ \footnote[3]{\texttt{g.tallarita@qmul.ac.uk}} 
}\\
\end{center}

{\it{ ${}^1$ %DESY, Notkestrasse 85, D-22603 Hamburg, Germany
Deutsches Elektronen-Synchrotron,\\ DESY, Notkestrasse 85, 22607 Hamburg, Germany
}\\
\vspace{0.5cm}
%\it{ ${}^2$ Queen Mary University of London\\ Centre for Research in String Theory \\ School of Physics and %Astronomy \\ Mile End Road, London, E1 4NS, UK}}

}

%%%%%%%%%%%%%%%%%%%%%%%%%%%%%%%%%%%%%%%%  Resets the counter to 0 and then uses Arabic numerals i.e. 1,2,3...
\setcounter{footnote}{0}
\renewcommand{\thefootnote}{\arabic{footnote}}
%%%%%%%%%%%%%%%%%%%%%%%%%%%%%%%%%%%%%%%%%%%%

\abstract{We extend the framework of general gauge mediation to cases where the mediating fields have a nontrivial spectral function, as might arise from strong dynamics.  Demonstrating through examples, this setup describes a broad class of possible models of gauge mediated supersymmetry breaking.  A main emphasis is to give general formulas for cross sections for $\sigma(visible\rightarrow hidden)$ in these resonance models. We will also give formulas for soft masses,  A-terms and demonstrate the framework with a holographic setup.}

%%%%%%%%%%% Introduction

\section{Introduction}
Within the paradigm of gauge mediated supersymmetry breaking, perturbative models of supersymmetry breaking and perturbative mediation have been well understood and well mapped out \cite{Giudice:1998bp}.   One of the key challenges, if supersymmetry breaking is also to be a solution of the hierarchy problem, is that supersymmetry is broken dynamically \cite{Witten:1981nf}, and hence the supersymmetry breaking sector is strongly coupled.

The formalism of general gauge mediation \cite{Meade:2008wd,Buican:2008ws,Marques:2009yu,Galli:2012jp} encodes the effects of supersymmetry breaking in a set of current correlators.  It assumes that when a set of parameters $(\{\alpha_{i}\},\{\lambda_j\})\rightarrow 0$, the supersymmetric standard model and the sector that breaks supersymmetry utterly decouple from each other. One can then work perturbatively in these parameters.  This framework is necessary to encode the effects of a breaking sector that is strongly coupled.  Further, through using duality transformations and techniques of effective field theory, it may be possible to find a perturbative description which would give the same structure for the current correlators, for instance.

Whilst the programme is quite general, it is often useful to classify models by making assumptions about the type of theory one is analysing.    In this paper we will assume that when the gauge fields of the standard model interact with the supersymmetry breaking sector their two point functions develop a non trivial spectral function.  In other words the assumption is made that vector mesons participate in the mediation. Whilst this is naturally the case in five dimensional models \cite{Mirabelli:1997aj,Abel:2010uw,McGarrie:2010yk,McGarrie:2011dc,McGarrie:2010kh,McGarrie:2010qr,Abel:2010vb,McGarrie:2011av}, and these will be used as examples, it is hoped that these results will have application for theories that are \emph{four} dimensional and \emph{strongly coupled}, such as in \cite{Abel:2012un}.

Taking our motivation from the hadronic world \footnote{recent developments in this direction may be found in \cite{Abel:2012un}.}, it is observed that at high energies photons behave like hadrons, except for their weaker coupling \cite{Bauer:1977iq,Gilman:1972yb,feynman1972photon}.  In particular there are intermediate resonances in photon-hadron interactions.  Building on this one may expect that when standard model gauge fields interact with a strongly coupled supersymmetry breaking sector it may interact via resonances.  One key focus will be to display this feature in scattering cross sections of \emph{visible}$\rightarrow$ \emph{hidden} processes.

Central to this point of view is the question of how one should think about the breaking sector when it is strongly coupled.  Our perspective is that  whilst one should expect messenger fields coupled to a supersymmetry breaking spurion \cite{Martin:1996zb} (which, in this analogy, we wish to loosely equate to the pion sector of QCD)  more fundamentally, from the perspective of gauge interactions, there will also be a vast array of resonances, many with the same quantum numbers as the standard model gauge fields.  This suggests not only new vector mesons in $U(1)_{em}$ but also color octets or ``colorons'' and ``colorinos'' of $SU(3)_c$ \cite{Kilic:2008pm,Kilic:2009mi} as the first signals of a supersymmetry breaking sector.

The resonances are bound states of hidden sector fields with the same quantum numbers as the standard model gauge fields. 
The resulting non trivial spectral function of the gauge fields will determine  the form factor.  The form factor, in this case, is a vector meson pole saturated charge distribution describing the interaction of external gauge fields with a strongly coupled sector.  

The form factor associated with the resonances encodes the low energy physics.  Additionally by constructing a tower of continually broadening resonances one may observe cross sections with a rather smooth scaling property \cite{Sakurai:1973rh} to higher energies.  

Whilst particular models will predict the form of the spectral function, ideally one would wish to fit the spectral function from observation,  through measuring cross sections \cite{Fortin:2011ad} such as those of  section \ref{crosssections}.  In this sense our approach will be mostly ``bottom-up'': we wish to extract soft masses, scattering cross sections and also to classify models from the behaviour of the spectral function associated with the resonances rather than specify a fully complete and perhaps overly complicated model from which one then attempts to derive these spectral functions.  At some points however, the AdS/CFT perspective is used in which one should associate the four dimensional form factor and spectral function as arising from a gauge field propagating in a higher dimensional space.

In a general way, this paper will give results for soft terms and scattering cross sections of visible matter to supersymmetry breaking fields.  Where it is useful, it will give examples of models that demonstrate key features of this rather broad class of model. A $U(1)$ toy model of gauge mediation is used in general, although these results might be applicable for all of the $SU(3)\times SU(2)\times U(1)$ gauge groups of the supersymmetric standard model.

Whilst completely compatible with four dimensional general gauge mediation, in addition the \emph{key} features are additional standard model charged resonances arising from the supersymmetry breaking sector and light sfermions relative to gauginos at the high scale $M$.
%In this type of mediation a typical feature is clear: the spectral function will have a threshold or resonance scale $\mu$, which is the lowest scale in which the two point function deviates from that of a free field (up to overall renormalisation Z).  The supersymmetry breaking sector has a characteristic scale $M$.  It is useful to therefore compare $M$ with $\mu$:  If $M<\mu$ the resonance scale will not appreciably effect observables.  If on the other hand on finds that $\mu \leq M$  observable quantities are suppressed by some ratio of these scales.  

\section{General Resonance Mediation}
In this section  the currents that encode supersymmetry breaking effects from fields of the supersymmetry breaking sector are introduced.  These current are extracted from a global symmetry of the supersymmetry breaking sector that is weakly gauged.
 
These currents will be used to derive general formulas for MSSM soft masses and also to compute scattering cross sections.
\subsection{The Currents}
 The supersymmetry breaking effects are encoded by introducing a real linear,
 \be
D^2 \mathcal{J}=\bar{D}^2 \mathcal{J}=0, 
\ee
global current supermultiplet $\mathcal{J}$
\be
\mathcal{J}=J+ i\theta j - i\bar{\theta}\bar{j}-\theta \sigma^{\mu}\bar{\theta}j_{\mu}+\frac{1}{2}\theta^2\bar{\theta}\bar{\sigma}^{\mu}\partial_{\mu}j-\bar{\theta}^2\theta\sigma^{\mu}\partial_{\mu}\bar{j}-\frac{1}{4}\theta^2\bar{\theta}^2\square J.
\ee
The currents are associated with a global symmetry of the supersymmetry breaking sector.  The current correlators of which  are defined in general gauge mediation \cite{Meade:2008wd} 
\be
i \tilde{C}_0 (p^2/M^2;M/\Lambda)=\int \frac{d^4x}{(2\pi)^4} \braket{J(x)J(y)}e^{ik.(x-y)}
\ee
\be
\!\!\!\!\!\!\!\!\!i \Pi_{\mu\nu}\tilde{C}_{1}(p^2/M^2;M/\Lambda)=\int \frac{d^4x}{(2\pi)^4} \braket{j_\mu(x)j_{\nu}(y)}e^{ik.(x-y)}
\ee
\be
\!\!\!\!\!\!\!\!\!\!\!\!\!\!\!\!\!\!\!\!i\sigma^{\mu}_{\alpha\dot{\alpha}}p_{\mu} \tilde{C}_{1/2}(p^2/M^2;M/\Lambda)=\int \frac{d^4x}{(2\pi)^4} \braket{j_\alpha(x)\bar{j}_{\dot{\alpha}}(y)}e^{ik.(x-y)}
\ee
\be
 \ \ \ \epsilon_{\alpha\beta }M\tilde{B}(p^2/M^2)=\int \frac{d^4x}{(2\pi)^4} \braket{j_\alpha(x)j_{\beta}(y)}e^{ik.(x-y)} \label{MBhalf}
\ee
with $\Pi_{\mu\nu}=(p^2\eta^{\mu\nu}-p^{\mu}p^{\nu})$. The scale $M$ is a typical mass scale of the hidden sector. The scale $\Lambda$ is a UV scale to regulate the integrals, which cancels in the ``supertraced'' combination that appears in scalar mass formulas:
\be
 \Omega\left(\frac{p^2}{M^2}\right)=[\tilde{C}_{0}\left(\frac{p^2}{M^2}\right)+3\tilde{C}_{1}\left(\frac{p^2}{M^2}\right)-4\tilde{C}_{1/2}\left(\frac{p^2}{M^2}\right)].\label{supertrace}
\ee
In analogy to the hadronic world, we are typically only interested in the ``hidden'' part of the full current correlator.

Weakly gauging a global symmetry of the supersymmetry breaking sector, it is associated with it a set of interactions
\be
   \mathcal{L}\supset  g' \int d^4 \theta  \mathcal{J}_{ \cancel{SUSY} } \mathcal{V}'. \label{source1}
\ee
This is a source term appearing in the generating functional.  In full generality one may also introduce a source term for the supersymmetric standard model gauge symmetry 
\be
      \mathcal{L}\supset  g \int d^4 \theta  \mathcal{J}_{SSM} \mathcal{V}.\label{source2}
\ee
Where the $\mathcal{V}$ and $\mathcal{V}'$ are  real $\mathcal{V}=\mathcal{V}^{\dagger}$ vector superfields which in components are given by 
\be
\mathcal{V}=-\theta \sigma^{\mu}\bar{\theta}A_{\mu}-i \theta^2 \bar{\theta}\bar{\lambda}+i \bar{\theta}^2\theta \lambda+ \frac{1}{2}\theta^2\bar{\theta}^2 D.
\ee
In complete generality the vector fields and couplings in \refe{source1} and \refe{source2} may differ in the bare action.   One only requires that for instance
\be
\frac{i\delta^2  \text{Log}  Z[\{ \mathcal{J}_i\} ] }{\delta j^{\mu}_{\cancel{SUSY}}(x) \delta j^{\nu}_{SSM}(y)}= \braket{A_{\mu}(x) A^{'}_{\nu}(y)}
\ee
defines a meaningful two point function.    

 In models for which one has a perturbative description, this spectral function arises quite often:  The spectral function encodes the overlap of the fields directly coupling to the supersymmetry breaking sector, with the gauge fields of  the supersymmetric standard model.  In Deconstructed models the spectral function arises in summing over mass eigenstates in the two point function of interaction eigenstates of the lattice  \cite{Son:2003et,McGarrie:2010qr}. In warped 5d models it coincides with the Kaluza-Klein states or resonances of the bulk theory \cite{McGarrie:2010yk} or when fields depend on an additional parameter  \cite{Sundrum:2011ic}.  Examples in the next sections will show that to model vector meson dominance and its generalisations requires that the supersymmetry breaking currents do not directly couple to external fields but instead to intermediate resonances in the weakly coupled dual description.

 In this paper it is assumed that it is a general consequence of the strong dynamics of the supersymmetry breaking sector. Within gauge mediation, one could hope that every mediating sector could be described in terms of a specified spectral function.  

Assuming that the two point function for the mediating gauge fields between each set of sources is given by a general K\"all\'en-Lehmann spectral representation,
\be
\int d^4 x  \ e^{ip.x}\braket{A_{\mu}(x) A^{'}_{\nu}(0)}=\int d\sigma^2 \frac{i \rho(\sigma)  \Pi^{\mu\nu}(p^2)}{p^2(p^2-\sigma+i\epsilon)}\label{spectral}
\ee
Following the Feynman prescription, which will  be dropped from the notation henceforth. The $\rho(\sigma^2)$ is a spectral function.  Multiplying by $p^2$, a convention has been chosen in which the function is dimensionless.  In this way one may derive general results which may be applicable to particular models by specifying the exact form the spectral function.  Additional poles in the spectral function may be thought of as composite particles as these do not have associated fields in the Lagrangian.  The use of effective actions will allow us to determine these poles as they have an associated field in the effective description.

If one considers the mediating fields to be described by the spectral representation \refe{spectral}, then the scalar soft masses are given by 
\be
m_{\phi}^2= -(4\pi\alpha)^2\! \! \! \int \! \! \frac{d^4 p}{(2\pi)^4} \frac{1}{p^2}\int\!\! d\sigma^2 \!\! \int \! d\sigma^{'2} \frac{\rho(\sigma^2)}{p^2-\sigma^2}\frac{\rho(\sigma^{'2})}{p^2-\sigma^{'2}}\Omega\left(\frac{p^2}{M^2}\right).\label{sfermionmasses}
\ee
to leading order in $\alpha_{SM}$. Assuming a $U(1)$ toy model, ignoring group factors and  taking $\rho_a=\rho$ for all $a=0,1/2,1$.  There is a typical mass scale of the theory $M$, which for perturbative models is the messenger mass scale. 

In \refe{sfermionmasses} two spectral integrals have been explicitly pulled out from inside the correlators, associated with the outer loop of the typical two-loop diagrams.  The expression above is equivalent to that of general gauge mediation \cite{Meade:2008wd} in the presence of a ``modified current operator" \cite{Kroll:1967it}. The outer loop will obtain corrections at higher order in $\alpha_{SM}$ \cite{Lee:2010kb}.   We wish to stress that in using the spectral function we are not simply referring to the sub leading corrections in $\alpha_{SM}$ that arise quite normally in perturbation theory, but are instead referring to the spectral function that may arise from interacting with the strongly coupled hidden sector.

Despite the appearance of a spectral function, one may still expect a decoupling of the visible and hidden sector as $\alpha_{SM}\rightarrow0$.  Additionally for perturbative theories one may expect to return  to a trivial spectral function 
\be
\lim_{s\rightarrow \infty}\rho(s)=s\delta(\sigma^2),
\ee
where $s=p^2$, for which one obtains the results of general gauge mediation \cite{Meade:2008wd}.

 The factorising form of the scalars masses is also observed in semi-direct gauge mediation \cite{Argurio:2009ge} and in this case may be thought of as descending from the factorised form of the generating functional that defines the model in question.  A similar result is demonstrated  for holographic mediation models in section \ref{holo}.

The gaugino masses will be given by  another current correlator \refe{MBhalf}
\be
m_{\lambda n,m}= (4\pi \alpha) M\tilde{B}_{1/2}(0) f_n f_m  \label{Gauginomass}
\ee
where $f_n$ are some functions determined by the particular model.  As the gaugino mass is evaluated at $p^2=0$  it is unnaffected by the form factors that effect the sfermion masses and cross sections.

As models of this type fit within the framework of general gauge mediation,  these models typically demonstrate the same features, that being flavour universality, compatibility with gauge unification, a gravitino or multiple pseudo-goldstini as LSPs with standard model superpartner NLSP and small A-terms.  There are also certain sfermion sum rules \cite{Meade:2008wd}.   In addition one expects light sfermion masses relative to gaugino masses at the high scale $M$.

%\subsection{Relation to QCD}
%We wish to emphasise that this feature has been extensively studied in low energy effective models of QCD \cite{Bando:1987br}.  Initially Vector Meson Dominance models where developed and then generalised.  In some sense they describe an emergent property or Hidden Local Symmetry \cite{Harada:2003jx} which have a contemporary description in terms of quivers.   $\text{HLS}_1$ is a two site quiver model with one emergent gauge group and therefore one vector meson.  N-site linear quiver models and in particular $N\rightarrow \infty $,    encapsulate the feature of a tower of resonances and furthermore approach a  continuum $5d$ limit. HLS models were considered to be dual to QCD \cite{Harada:2003jx} directly in the sense of Seiberg duality \cite{Seiberg:1994pq}.   Recent emphasis has been made of the precise relation of Seiberg Duality and  Supersymmetric HLS \cite{Komargodski:2010mc,Abel:2012un}.   Significant overlap is made of the  $\text{HLS}_{N} /QCD$  and $AdS_5/QCD$ type constructions.    This is the context in which we develop these results.

\subsection{Minimal Messenger Sector}
Throughout this paper we will display both general results for an unspecified supersymmetry breaking sector and additional results based on the minimal messenger sector, which we use as a test model.  The superpotential for the minimal messenger sector is given by
\be
W=X\Phi\tilde{\Phi}.
\ee
$X$ is a spurion or goldstino multiplet $X=M+\theta^2F$ and $\Phi,\tilde{\Phi}$ are messenger fields charged under a global symmetry which we associated with \refe{source1}.  The mass eigenstates are two complex scalars $\phi_+,\phi_-$ with masses $M^2\pm F$ and two 2-component fermions $\psi,\tilde{\psi}$ with mass $M$.  For clarity, the hidden sector is comprised of $X,\Phi,\tilde{\Phi}$ for which the fields $\Phi,\tilde{\Phi}$ are messengers.  The messenger fields should not be confused with the mediating sector, which are the gauge fields and resonances.

In such a model it is straightforward to determine the gaugino mass from \refe{Gauginomass}
\be
m_{\lambda}=2\frac{\alpha}{4\pi}\frac{F}{M} g(x)
\ee
where for $x=F/M^2$ where  for $x<1$,  $g(x)\sim 1$. 

It will also be useful in this paper to evaluate the ``super-traced" combination of current correlators  \refe{supertrace} in the limit $p^2/M^2\rightarrow 0$  found in \cite{McGarrie:2010kh}
\be
\lim_{p^2\rightarrow 0}\Omega(p^2/M^2)=\frac{-1}{(4\pi)^2}\frac{2}{3}x^2 h(x)
\ee
where $h(x)\sim 1$ for $x<1$.  It is independent of $p$ and will allow one to carry out the momentum integral on the form factors alone.

\section{Currents and Fields}
In this section we wish to discuss some features relating vector meson dominance model and currents, which will be useful for understanding the relationship between different examples and in particular the relationship between VDM and $AdS_5$ models.

When one has a perturbative description and a lagrangian one may extract currents, for instance $j^{\mu}(x)=\bar{q}\gamma^{\mu}q(x)$, from Noether's theorem.  The currents obey a current algebra and this arises from the commutators  or anticommutators of the fields $\bar{q}, q$ in the current.  Without a lagrangian one may apply this procedure in reverse and write a set of equal time commutators and Schwinger terms for some currents and construct a ``field current identity'' \cite{Lee:1967iu,Kroll:1967it,Kang:1969ki,sakurai1969currents}:
\be
j^{a}_{\mu}(x)=-C\rho_{\mu}^{a}(x)
\ee
where $\rho_{\mu}$ is a general vector field which we will later associate with a massive vector meson.  $\alpha$ is a group index and C is some so far undetermined coefficient. 
In fact this C may be determined \cite{Lee:1967iv}  $C=m^2_{\rho}/g_{\rho}$ from a dispersion relation for the form factor $F(q^2)$ where $F(0)=1$.  The field current identity is possible whenever $\rho_{\mu}$ arises from a gauge field of a hidden local symmetry \cite{Bando:1984ej}. In addition we should make some criteria:
Normally \emph{Universality} of the coupling $g_{\rho}$ to all hadronic currents arises as a result of current conservation \cite{GellMann:1964nj}.  \emph{Universality} should be implemented in the interaction basis which may differ substantially from the mass basis.   For a small but explicit breaking of the symmetry $b \mathcal{L}_{B}$ the charges of the symmetry will still relate physical states that are \emph{dominated} by the state corresponding to exact preservation of this symmetry. 

In a supersymmetric theory we may
%A dispersion relation is the statement
%\be
%F(q^2)=\int \frac{\rho(k'^2)dk'^2}{k'^2-k^2}=\frac{ag f}{m_\rho^2-k^2}+\int \frac{\rho'(k'^2)dk'^2}{k'^2-k^2}
%\ee
%which when evaluated at $k^2=0$ gives a sum rule as $F(0)=1$.
 extend this to  a ``supercurrent field identity" which is a real linear multiplet
\be
\mathcal{J}_{\rho}=- \frac{m_{\rho}^2}{g_{\rho}}\rho  \ \ \  \text{where}   \ \ \ D^2\rho=\bar{D}^2\rho=0, \label{CFI}
\ee 
 which we may couple to 
\be
g_{SM}\int d^4 \theta \mathcal{J}_{\rho}V_{external}.\label{CFI2}
\ee 
If one is to interpret the vector meson as a gauge field of a local symmetry, we obtain a notion of a hidden local symmetry that is emergent.   This notion of dominance may be generalised in which the form factor is saturated by a tower of meson poles.  This has the appearance of a two point function of Kaluza-Klein modes of a bulk to boundary propagator:
\be
F(q^2)\sim\sum_n \frac{m^2_n}{g_{v}}\frac{ g_{n \phi\tilde{\phi}}}{q^2-m_n^2}\label{FORM}
\ee
and provide some generalised dispersion sum rules from  $F(0)=1$.  To write a gauge invariant and Lorentz invariant effective lagrangian to describe the features of a tower of massive spin one fields,  requires a Higgs mechanism to generate masses for the gauge fields. Choosing then a quiver or extra dimension\footnote{Kaluza-Klein modes in $\mathcal{N}=1$ 5d SYM also receive their masses from a supersymmetric Higgs mechanism.} to generate the tower of states.  These features may be generalised to $N$ local symmetries: $HLS_N$.  Mediation between two boundaries on a ``slice of AdS'' generalises this notion to a continuum.  The corresponding fields of $\mathcal{N}=1$ SYM in 5d that  fill the supercurrent field identity are 
\be
\mathcal{J}_{\rho} \supset  ( \   A^{\mu} ,\   \lambda^{\alpha}, \  \bar{\lambda}^{\dot{\alpha}}, \  D_{5}\Sigma ).
\ee
  As $\mathcal{J}_{\rho}$ is a set of operators, the field current identity is an operator-field correspondence from the perspective of AdS/CFT.

\section{Examples}
It is useful to compare some typical  examples of models that exhibit the features of resonance mediation. In this section we will complete a (not exhaustive) survey of different concrete models which realise such a scenario.

\subsection{Vector Meson Dominance/Two Site Quiver}
Perhaps the simplest and most well motivated model is of vector meson dominance \cite{Nambu:1997vw,GellMann:1961tg,Nambu:1962zz}.  Within this class of model are many variants,  depending on the number of vector mesons and also whether the standard model gauge fields couple directly to the hidden sector fields or whether they must couple through a vector meson.  We will take as our benchmark the two site quiver model  \cite{McGarrie:2010qr,McGarrie:2011dc,Sudano:2010vt,Auzzi:2010mb}, as in figure \ref{diagram}, as the supersymmetric action is well defined.  Considering one vector meson with mass $m_v$.  It is also worthwhile to consider how this feature arises in the magnetic description of SQCD \cite{Abel:2012un}.

The spectral function may be written to take the form (see also section \ref{A Toy Model})
\be
\rho = m^2_v\delta (\sigma^2-m^2_v) 
\ee
or equally 
\be
\rho = s\delta(\sigma^2) - s\delta (\sigma^2-m^2_v) 
\ee 
using a manipulation similar to \refe{manip},
resulting in an sfermion mass formula 
\be
m_{\phi}^2= -(4\pi\alpha)^2\! \! \! \int \! \! \frac{d^4 p}{(2\pi)^4} \frac{1}{p^2}\left[\frac{m^2_v}{p^2+m_v^2}\right]^2\Omega\left(\frac{p^2}{M^2}\right)\label{sfermionvmd}
\ee
after Wick rotation.    These models have been explored and generalised \cite{McGarrie:2010qr,McGarrie:2011dc,Auzzi:2010mb,Auzzi:2011eu,Auzzi:2011wt,Auzzi:2011gh,Auzzi:2010xc}.
Typically one finds three regimes: a) $m_v\ll M$ where there are additional contributions to sfermions masses at three loop \cite{McGarrie:2011dc} and the model is ``Gaugino Mediated".  Conversely in  the regime  b) $m_v\gg M$ the form factor 
\be
F(p^2)=\left[\frac{m_v^2}{p^2+m_v^2}\right] \label{formfactor}
\ee
is $F\simeq 1$. In this case the standard model gauge interactions may be considered to be directly coupled with the supersymmetry breaking sector and not through an intermediate resonance. This will give soft masses of \cite{Martin:1996zb}. The hybrid regime c) in which $m_v\sim M$ typically has suppressed but non zero scalar soft masses at the high scale and essentially MSSM RG running from the scale $M$ down.

In section \ref{A Toy Model} we will derive cross sections of $\sigma(visible\rightarrow hidden)$  that reproduce the structure of $\sigma(e^{+}e^{-}\rightarrow hadrons)$ as found within the vector meson dominance model.

For a generalised vector meson model the more canonical form of the spectral function is
\be
\rho(\sigma^2)=\sum_v \frac{m_v^2}{g_v}\delta(\sigma^2-m_{v}^2)\label{canonical}.
\ee

It is interesting to explore adding to the $\rho(\sigma) = resonances+ 
continuum $ where the continuum piece can take the form of 
\be
\rho_{cont.} = R(s)_{pert.} \theta(\sigma^2-s_c).
\ee
where $R(s)_{pert.}$ may be chosen to match the perturbative (S)QCD description \cite{Shifman:1978by}.   Seiberg dual models \cite{Green:2010ww},  as they have a single vector meson, are an ideal benchmark construction to test scattering cross sections both at weak and strong coupling in both the electric and dual magnetic descriptions.

\subsubsection{Old  Vector Meson Dominance versus Quivers}
The original proposals of vector meson dominance require that the standard model gauge fields interact  with hadronic currents via an intermediate vector meson.  Generalising this to supersymmetry one may build current vector couplings
\be
\mathcal{L}\supset \int d^4 \theta \  \left[g_{SM} \mathcal{J}_{SSM}\mathcal{V}_{SSM}     +  \  g_{SM} \mathcal{J}_{\rho}\mathcal{V}_{SSM} +  \  g_{\rho} \mathcal{J}_{\cancel{SUSY}}\mathcal{V}_{\rho}\right]
\ee
 and implement an action that couples the standard model gauge field to the vector mesons such as \refe{CFI2}.  Such a model would generate soft masses for the standard model sfermions at the two loop level and would generate a mass for $\lambda_{\rho}$
\be
g_{\rho}^2 M\tilde{B}_{1/2}(0)\lambda_{\rho}\lambda_{\rho}.
\ee
The standard model gaugino $\lambda_{SM}$  would however remain massless! 

Instead if the vector mesons are interpreted as mass eigenstates  $\tilde{V}_i$ and the currents couple to interaction eigenstates $V_j$ with $i=0,1$, as in the quiver models  \cite{McGarrie:2010qr,McGarrie:2011dc,Auzzi:2010mb,Auzzi:2011eu,Auzzi:2011wt,Auzzi:2011gh,Auzzi:2010xc}, the currents couplings are 
\be
\mathcal{L}\supset \int d^4 \theta \  g_{0} \mathcal{J}_{SSM}\mathcal{V}_{0}     +   \int d^4 \theta \  g_{1} \mathcal{J}_{\cancel{SUSY}}\mathcal{V}_{1}.
\ee
In such a scenario soft masses are generated as in \refe{sfermionvmd} for the sfermions and also
\be
g^2_{SM}M\tilde{B}_{1/2}(0)\tilde{\lambda}_{i}\tilde{\lambda}_{j}
\ee
gaugino masses are generated for both $\tilde{\lambda}_{j}$ mass eigenstates\footnote{One must also take into consideration the Dirac mass that arises from the action of the quiver.}. 
   In addition the quiver model or hidden local symmetry model interprets the vector meson as a gauge boson of an emergent gauge symmetry.  It naturally incorporates a gauge invariant (and supersymmetric) Higgs mechanism to generate meson masses. 
%\subsubsection{IR Plus UV Equals Full}
%In principle we wish to fit the form factor for both large and small $Q^2$.   
%\begin{description}
%\item[$Q^2\ll m^2_v$] 
 %In this regime the form factor is given by 
%\be
%F(-Q^2)=[\frac{m_v^2}{Q^2+m_v^2} ] \label{formfactor}
%\ee
%\item[$Q^2\gg m^2_v$]  Naively these models $F\simeq 1 $ in this limit.  If we have crossed over $\Lambda$ the cutoff of the effective description then we may expect some additional features.

% Using the Operator Product Expansion we could try and fit
%\be
%F(-Q^2)=\frac{16\pi f^2_{\phi}\alpha(Q^2)}{Q^2}[1+C_0\frac{\alpha(Q^2)}{\pi}+\sum_n \frac{C_{2n}}{Q^{2n}}]
%\ee%We hope that overlap can be made between this form factor programme and the OPE programme %\cite{Fortin:2011ad}.
%\end{description}
%It would be interesting to understand if there are Instanton contributions to the form factor \cite{Faccioli:2002jd}.

\subsection{A Flat Extra Dimension}
A well known example is the flat extra dimension\cite{Mirabelli:1997aj,McGarrie:2010kh}.  In these models the Lagrangian is $\mathcal{N}=1$ Super Yang Mills in five dimensions on an $R^{1,3}\times S^1/\mathbb{Z}_{2}$ background.  The supersymmetric standard model is placed at a fixed point $x_5=0$ and a supersymmetry breaking sector located at $x_5=\ell$, where $\ell$ is the length of the fifth dimensional interval. In this model the spectral function takes the form
\be
\rho (\sigma)= -\sum_n p^2 (-1)^n \delta (\sigma^2-m_n^2) = \sum_n m^2_n (-1)^n \delta (\sigma^2-m_n^2) \label{specfunction}
\ee
where $m_n=n\pi/\ell$.   In the above we have used the identity
\be
\delta(0)=\frac{1}{2\ell}\sum_n\frac{p^2-m^2_n}{p^2-m^2_n}.\label{manip}
\ee
The current couplings take the form 
\be
\mathcal{L}\supset\int d^4 \theta \  g \mathcal{J}_{SSM}\mathcal{V} \delta(y-0)    +   \int d^4 \theta \  g \mathcal{J}_{\cancel{SUSY}}\mathcal{V} \delta(y-\ell)
\ee
where  $g_{SSM}=g_{\cancel{SUSY}}=g$: as the interaction eigenstates are mass eigenstates \emph{universality} of couplings is implemented automatically.
The spectral function  determines a leading order sfermion soft mass formula
\be
m_{\phi}^2= -(4\pi\alpha)^2\sum_{n,n'} \int \! \! \frac{d^4 p}{(2\pi)^4} \frac{(-1)^{n+n'}}{p^2-m^2_n}\frac{p^2}{p^2-m_{n'}^{2}}\Omega\left(\frac{p^2}{M^2}\right).
\ee
Completing a Matsubara summation the form factor is given by 
\be
F(p\ell)= \left(\frac{p\ell}{\sinh{p\ell}}\right).
\ee
The gauge couplings for such a construction  are UV sensitive and display a power law running \cite{Dienes:1998vg}.  The key features of the spectral function are that the squared masses scale with $n^2$ and that the coefficient oscillates sign, which provides a useful cancellation also for pion form factors \cite{RuizArriola:2008sq}, a feature that may be modelled with brane to bulk wavefunctions as 
\be
 f_n(x_5)f_n(y_5)=(e^{ik_5.(x_5-y_5)}+e^{ik_5.(x_5+y_5)}).
\ee
The oscillating sign has been noted in Large $N_c$ form factors also \cite{Dominguez:2010rp}.  These models share the same three regimes as the quiver models.

\subsection{General Regge-like Mediation}
Motivated by low energy models of QCD  we could assume  a Regge-like trajectory \cite{Collins:1977jy} which scales with $n$ instead of $n^2$
\be
m_n^2=\mu^2(n+S)  \ \ \text{with} \ \ \  \mu^2= \frac{2}{\alpha'}
\ee
with $\alpha'$ associated to the slope of the Regge trajectory, $\mu^2$ being the confining string tension. The full Regge trajectory is associated with spin $S$ however we will simply focus on the vector mesons
where we take
\be
\rho(\sigma)=  \sum_n p^2 F^2_n (-1)^n  \delta (\sigma^2-m^2_n)=-\sum_n m_n^2 F^2_n (-1)^n  \delta (\sigma^2-m^2_n).
\ee
We have assumed the $(-1)^n$ factor for suitable convergence of the Matsubara summation.

Using the above spectral function will result in
\be
m_{\phi}^2=-(4\pi\alpha)^2\sum_{n,n'} \int \! \! \frac{d^4 p}{(2\pi)^4} \frac{F^2_n(-1)^{n+n'}}{p^2-\mu^2n}\frac{F^2_{n'}p^2}{p^2-\mu^2 n'}\Omega\left(\frac{p^2}{M^2}\right).
\ee
For constant $F_n=F$ we may complete a  Matsubara summation
\be
\sum_{n=-\infty}^{+\infty} \frac{(-1)^n}{p^2-\mu^2 n}=  \frac{\pi}{\mu^2} \text{cosec}\left(\frac{\pi p}{\mu^2}\right).
\ee
Using the above identity, unfortunately the relevant momentum integral does not appear to converge:  we have in mind a more complete picture \cite{Dominguez:2004bx,RuizArriola:2008sq}.      For instance, ideally one would wish to be able to derive  a form factor as a ratio of Gamma functions, inspired by the Veneziano amplitude, as was suggested for pions:
\be
F(s)\sim\frac{\Gamma(1-\alpha(s))}{\Gamma(\lambda-\alpha(s))}\frac{\Gamma(\lambda-\frac{1}{2})}{\Gamma(\frac{1}{2})},
\ee
where $2<\lambda<2.5$ fit the pion data quite well \cite{cumming1971hadronic}.  This scales as $F(s)\sim 1/s^{\lambda-1}$ in the UV.  For the case of pions the above result for a $1\rightarrow 2$ amplitude was obtained from using certain current algebra techniques on a $2\rightarrow 2$ amplitude and it would be desirable to derive this type of result more directly for this supersymmetric case. 

However for $\mu \sim M$ this model will have similar features to that of the two site quiver model (see for instance  \cite{Auzzi:2010xc}) with only one resonance.   For a minimal messenger sector \refe{superpotential}, the typical scalar mass result is then
\be
m_{\phi}^2\sim \left(\frac{\alpha}{4\pi}\right)^2 \left|\frac{F}{M}\right|^2\left|\frac{\mu}{M}\right|^{\tau}  \ \ \ \text{where} \ \  \tau \in [0,2]
\ee
with gaugino soft masses unchanged.  Surprisingly little has been said about this model in regard to supersymmetry breaking.

\subsubsection{From A Soft Wall Model}
A model that scales with $n$ instead of $n^2$ may be constructed \cite{Karch:2006pv,Abidin:2009hr} using the AdS/CFT perspective.   To construct this model one couples $\mathcal{N}=1$ Super Yang Mills in 5d to a Dilaton multiplet
\be
S= \int d^5 x \sqrt{g}e^{-\Phi}\mathcal{L}_{SYM}
\ee
whose scalar profile
is given by 
\be
\Phi(z)=a z^2
\ee
where $a$ is a positive constant.  We choose the Poincar\'e metric:
\be
ds^2= e^{2A(z)} (\eta^{\mu\nu}dx_{\mu}dx_{\nu}-dz^2 ) \ \ \ \text{with}  \ \ \   A(z)=-\text{Log} (z),
\ee
such that $G_{MN}=\eta_{MN}/z^2$, where $z\in [z_0,\infty)$.
In this case we can introduce a supersymmetry breaking sector located at some peak value of $z=L_{1}$ perhaps with a delta function 
\be
   S\supset    \int d^5x  g' \int d^4 \theta  \mathcal{J}_{\cancel{SUSY}} \mathcal{V}'\delta(z-L_1)  \label{source3}
\ee
 or with some smoother profile. This would generate an effective action  located around $z=L_1$.  Exact locality in position space means complete delocalisation in momentum space and it is therefore likely that a smoother profile will dampen couplings to modes with very different Kaluza-Klein number to the incoming mode.

\subsection{The Unparticle Limit}
One could imagine an approximately continuous set of resonances \cite{Georgi:2007ek,Stephanov:2007ry,Cacciapaglia:2008ns} above some scale $\mu$, which would arise from coupling the standard model to an approximately conformal sector, that mediate the supersymmetry breaking effects.

Let us attempt a crude estimate of the soft terms of such a model.  We will not compute the full two loop diagrams but suppose that due to  the constraining effects of supersymmetry\footnote{A book-keeping construction analogous to ``theta-warping'' may be fruitful here \cite{Bagger:2011na}.}, one may use the scalar unparticle two point function on the outer loop for all diagrams: For the ungauge boson, ungaugino and dynamical unD-term, $D_5\Sigma$.

The scalar two point function is given by
\be
\Delta(p,\mu,d_s)=i \frac{A_{d_s}}{2\pi}\int^{\infty}_{\mu^2}(M^2-\mu^2)^{d_s-1}\frac{1}{p^2-M^2+i\epsilon} dM^2 \nonumber
\ee
\be
=i \frac{A_{d_s}}{2\sin d_s \pi}(\mu^2-p^2-i\epsilon)^{d_s-2}+...
\ee
where \cite{Georgi:2007ek}
\be
A_{d_s}=\frac{16\pi^{5/2}}{(2\pi)^{2d_s}}\frac{\Gamma (d_s+1/2)}{\Gamma(d_s-1)\Gamma(2d_s)}.
\ee
$d_s$ is the scaling dimension.  This unparticle description is meant to  overlap with that of the holographic models, where in both cases the IR scale or IR brane implement the breaking of conformal symmetry. In the case of an IR cutoff one obtains a mass gap and then a discrete spectrum of massive states.   Alternatively a $z$ dependent Dilaton profile coupled to the bulk SYM action \cite{Marti:2001iw} may be used to implement a soft breaking of conformality and a continuum spectrum of states. 

In the limit that  $\mu\ll M$ with a minimal messenger sectior $W=X\Phi \tilde{\Phi}$  the super-tracted combination \refe{supertrace} of current correlators 
should be expanded in $p^2/M^2$ and the leading term is independent of $p$ \cite{McGarrie:2010kh}.  Therefore we need only evaluate the outer loop 
\be
\int \frac{d^d y}{(2\pi)^d} \frac{y^2}{(y^2+\Delta)^{\alpha}}=\frac{1}{(4\pi)^{d/2}}\frac{d}{2}\frac{\Gamma(\alpha-d/2-1)}{\Gamma(\alpha)} \Delta^{1+d/2-\alpha}
\ee
with $\alpha=4-2d_s$.   %  We obtain 
%\be
%\mu^{-2+4d_s}\frac{(-1)^{2\alpha-3}}{(4\pi)^2}\frac{\Gamma(\alpha)}{(\alpha-3)!}[2(\gamma + 1 + ...+ \frac{1}{3-\alpha})+2\ln\frac{4\pi}{\mu^2}+1+\frac{2}{\epsilon} ]
%\ee
To couple the ungauge fields to the breaking sector we suggest an operator of the form
\be
S\supset \int d^4x \int d^4\theta \frac{g'}{\Lambda^{d_s-1}}\mathcal{ J} \mathcal{V}^{d_s}_{unparticle}
\ee
where for simplicity $\Lambda=M$.  This allows one to repackage $\mu/M$ appropriately and the sfermion masses scale as
\be
m_{\phi}^2\sim \left(\frac{\alpha}{4\pi}\right)^2\left(  \frac{A_{d_s}}{2\sin d_s \pi}\right)^2\frac{1}{\Gamma(4-2d_s)} \left|\frac{F}{M}\right|^2\left|\frac{\mu}{M}\right|^{4d_s-2}.
\ee
Higher order supersymmetry contributions will certainly effect ungauginos, but as the ungaugino soft mass operator couples to all the ungauginos, it is not entirely clear how resolved individual resonances may become.   This is left open for a more systematic study.

\section{Scattering: A Toy Model}\label{A Toy Model}
We now turn to scattering which are the main results of this paper. 
In this section we wish to develop some intuition relating low energy pion and vector meson physics with its SQCD and supersymmetry breaking analogues.  Building on this intuition one will be able to extract interesting results for more general situations.

%%%%%%%%%%%%%%%%%%%%%%%%%%%%%%%%%%%%%%%%%%%%%%%%%%%%%%%%%%%%%%%%%%%%%%%%%%%%%%%%%%%%%%%%%%%%%%%%%%%%%%%%%%%%%%%%%%%%%%%%%%%%%%%%
\begin{figure}[ht]
\begin{center}
\includegraphics[scale=0.8]{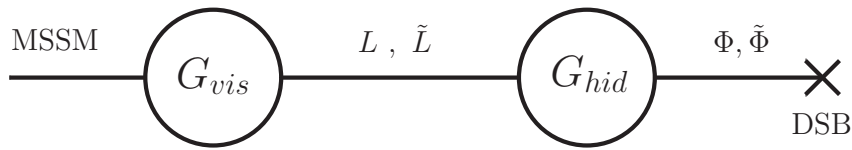}
\caption{The two site quiver as a toy model.}
\label{diagram}
\end{center}
\end{figure}
%%%%%%%%%%%%%%%%%%%%%%%%%%%%%%%%%%%%%%%%%%%%%%%

The toy model is the supersymmetric two site quiver \cite{McGarrie:2011dc,McGarrie:2010qr,Auzzi:2010mb,Auzzi:2010xc,Green:2010ww}. There is a vector superfield for each gauge group $V_{i}$ where $i=0,1$ labelling the visible and hidden sector respectively. Taking a simple hidden sector superpotential
\be
W=X\Phi\tilde{\Phi}+ K(L\tilde{L}-v^2)  \label{superpotential}
\ee
The supersymmetry breaking spurion $X=M+\theta^2F$ couples to messengers $\Phi,\tilde{\Phi}$.  The field $L,\tilde{L}$ are linking fields and the field $K$ is a Lagrange multiplier superfield.

The scalar messenger field mass eigenstates $\phi_+,\phi_-$ have mass squareds $M^2\pm F$, the fermions $\psi, \tilde{\psi}$ have mass $M$.   It is useful in this analogy to imagine these states as like pions of QCD as they will interact with vector mesons and are ultimately composite, although appear fundamental.  Additionally, when we evaluate the correlators their average mass will define the value of the branch cut $s_0=4M^2$.  The analogy breaks down as the pions are part of an $SU(2)$ adjoint representation and we are taking the messengers to be in the fundamental.

In the interaction basis  there are two ($i=0,1$) gauge fields $A^{\mu}_{i}$ associated with the visible and hidden sector lattice sites.  We will label the external gauge field that couples to the supersymmetric standard model as $A^{\mu}_{0}=\gamma^{\mu}$.   In the mass basis we will label $\tilde{A}^{\mu}_i$ and in particular  the massive state $\rho^{\mu}=\tilde{A}^{\mu}_{1}$, with masses $m_0=0$ and $m_1=m_{\rho}$. Similarly for the states $\tilde{\lambda}_1$and $\tilde{\lambda}^{\alpha}_2=\rho^{\alpha}$  .

In these two site models 
\be
\frac{1}{g^2_{SM}}=\frac{1}{g^2_0}+\frac{1}{g^2_1}   \ \ \   \ g^2_{SM}=\frac{g^2_0 g^2_1}{g^2_0+g^2_1} \label{gcouplings}
\ee
where 
\be
m_v^2=4v^2 (g^2_0+g^2_1)= 4 v^2\frac{ g^2_0g^2_1}{g^2_{SM}}.
\ee
One can identify the mass eigenstates
\be
 \rho_{\mu}=\frac{g_0 \gamma_{\mu}-g_1A^1_{\mu}}{\sqrt{g^2_0+g^2_1}}  \ \  \text{and}  \ \  A_{\mu}=\frac{g_1 \gamma_{\mu}+g_0A^1_{\mu}}{\sqrt{g^2_0+g^2_1}} .
\ee
In the scattering process of $\gamma^{\mu}$ to $\phi_+,\phi_-$ there is an intermediate  $\rho^{\mu}$ as a mass eigenstate.  
Crucially, the  form factor for  $A\rightarrow B+\gamma$ for A, B being hidden sector fields, may be related to the form factor of the process $A\rightarrow B+A^{\mu}_1$
\be
F_{AB}^{\gamma}(q^2)=\left[\frac{m_{\rho}^2}{m_\rho^2+q^2}\right]F_{AB}^{A^1_{\mu}}(q^2)=-\sum_{n=0}^1 \frac{(-1)^n}{q^2+m_n^2} q^2 F_{AB}^{A^1_{\mu}}(q^2)
\ee
where on the right hand side we have used the ``bulk propagator" which is derived from the overlap of the interaction basis with the mass eigenstates \cite{Schildknecht:1972ej}.  One is therefore able to define a ``modified current operator'' \cite{Kroll:1967it}
%\cite{Cornwall:1971pk,Gilman:1972yb,Dominguez:1982xb}
\be
\braket{B|\hat{J}^{\gamma, \mu}|A}= D^{\mu\nu}(q^2) q^2 \braket{B|J^{A^1_{\nu}}_{\nu}|A}
\ee
where a given amplitude $\mathcal{M}$ is found from the Heisenberg field $j^{\mu}$, 
\be
\mathcal{M}^{\mu}(p)= \int d^4x e^{ip.x}\braket{B|j^{\mu}(x)|A}.
\ee
One takes the absolute square of this matrix element to compute cross sections.   Such an identity is crucial in the vector meson dominance program in $e^{+},e^{-}\rightarrow hadrons$ and similarly for $e^{+}e^{-}\rightarrow \mu^{+}\mu^{-}$ with intermediate $\phi(1020)$ and also $\omega(782)$ decays \cite{Nambu:1962zz}.  Here we predict the same feature in  $e^{+}e^{-}\rightarrow hidden$.  
%$m_{\rho}^2= ag^2_{\rho\phi\phi}f^2_{\phi}=g_{\rho\phi\phi}g_{\rho\gamma}$.

It would therefore be possible to measure $g_0$ and the combination $g^2_{SM}$ independently.  This is done by measuring\footnote{This may take different forms depending on where one interprets $g_0=g_{SM}$ or as in \refe{gcouplings}.}
\be
\frac{g^2_{0}}{4g^2_{SM}}=\frac{1}{g_0^2}\int_{s_0} d s' \sigma_{a}(visible\rightarrow \rho \rightarrow \phi_+,\phi_-s')
\ee
where the cross section $\sigma$ is related to the current correlators $\tilde{C}_a(s)$ \cite{Bauer:1977iq}.  Measurements of this type would be a direct test of the magnetic gauge coupling $g_1$ in Seiberg dual models of the form \cite{Green:2010ww,Komargodski:2010mc}.  It is also an example of an integral of a cross section. Let us now demonstrate a sum rule  from another intergral over a cross section.

\subsection{A Sum Rule}
It is interesting to  also explore possible sum rules that may arise.
For the case of a single vector meson as in the above model, it is possible to construct a sum rule for the cross sections that will apply for all $\tilde{C}_a(s)$ separately.   Defining
\be
\Sigma^{vis}_{\rho\rho}(m^2_{\rho})=m_{\rho}\Gamma(\rho\rightarrow vis) \ \  \text{and}   \  \   \Sigma^{hid}_{\rho\rho,a}(m^2_{\rho})=m_{\rho}\Gamma_a(\rho\rightarrow hid)
\ee
 ($\Sigma^{vis}_{\rho\rho}(m^2_{\rho}$) carries a factor $4\pi \alpha^2$) where in general one has 
\be
\Sigma^h_{V_1V_2}=\frac{1}{6}S_F(2\pi)^4 \delta_a(p-p_F)\braket{h|J^{V_1}_{\mu}(0)|0 }\braket{h|J^{V_2}_{\nu}(0)|0 }^*\eta^{\mu\nu}
\ee
where $\bra{h}$ labels hidden sector states and $S_F$ being the standard phase space integration.  Redefining the form factor of the modified current operator to be 
\be
F(s)=\left[\frac{-g_e m_\rho^2 }{g_{\rho}}\frac{ g_{\rho\phi\tilde{\phi}}  }{s-D(s)}\right]
\ee
 one can write
\be
\sigma_a(vis\rightarrow hid)=\frac{12\pi}{s}\frac{\Sigma^{vis}_{\rho\rho}(s)\Sigma^{hid}_{\rho\rho,a}(s)}{|s-D(s)|^2},
\ee
where $D(s)=m_{\rho}^2+\Pi(s)$, the last term being the self energy. Using the dispersion relation
\be
\frac{1}{D(s)-s}=\frac{1}{\pi}\int_{s_0}^{\infty}\frac{\text{Im} D(s')}{|D(s')-s'|^2}\frac{ds'}{s'-s+i\epsilon}
\ee
with  $D(m_{\rho})=m_{\rho}^2-im_{\rho}\Gamma$, the sum rule is
\be
\frac{m^2_{\rho}}{D_b(0)}=\frac{g_{\rho\phi\tilde{\phi}}}{g_{\rho}}=\frac{1}{12\pi^2 m_{\rho} }\sum_{cuts}\int^{\infty}_{s_0} s \sigma_b(vis\rightarrow \rho,s)ds /{\Gamma(\rho \rightarrow vis)}= x_{b,\rho}
\ee
This is a correction parameter due to the finite width resonance.  In the zero width approximation $D(0)=m_{\rho}^2$ which leads to  $g_{\rho\phi\tilde{\phi}}=g_{\rho}$ in \refe{FORM}.  Taking $D(0)=a_{\rho}m^2_{\rho}$ at finite width and   measuring $x_{\rho}$ in conjunction with $m_{\rho}$ can be used to measure the value of $a$ in \cite{Green:2010ww,Komargodski:2010mc,Abel:2012un}.  If $x_{b,\rho}=x_{\rho}$ for all correlators, implying a new supersymmetric type of universality of form factors, this would be a highly nontrivial check on the emergence of a  hidden local symmetry.   It can be shown that the sum of  $\sum_v x_v  m_v^2/g_v=C_{Sch}$ defines the Schwinger term in equal time commutators. There may be further sum rules based on already known QCD sum rules such as this.

It is also natural to define the messenger's effective charge radius 
\be
\braket{r^2}=-6\frac{\partial}{\partial s}F(s)|_{s=0}
%= \frac{1}{\pi}\int_{s_0}^{\infty}\frac{\text{Im}F(s)}{s^2}ds
\ee
which for the two site model above is given by 
\be
\braket{r^2}= \frac{6}{m_{\rho}^2}.\label{r}
\ee
In summary, for this simple two site quiver the variables $g_1$, $m_{\rho}$ and $x_{\rho}$ (and therefore $a_{\rho}$)  may all be determined from measuring scattering cross sections of visible $\rightarrow$ hidden sector states, which is directly related to the proposals of \cite{Green:2010ww,Komargodski:2010mc,Abel:2012un}.
\section{Scattering Cross Sections}\label{crosssections}
In \cite{Fortin:2011ad} cross sections\footnote{See also pages 242 e.q. (5-155a)  and page 310 of \cite{Itzykson:1980rh}.} from hidden sector current correlators were computed.
 This section  explores the effect of intermediate states in these cross sections of visible+visible $\rightarrow$ hidden processes. It is  hoped that these results will give some insight into connecting the theoretical models with observation.  In particular the approach taken here is form factor based and it would be interesting to develop methods to overlap this with the OPE approach of \cite{Fortin:2011ad}.
%%%%%%%%%%%%%%%%%%%%%%%%%%%%%%%%%%%%%%%%%%%%%%%%%%%%%%%%%%%%%%%%%%%%%%%%%%%%%%%%%%%%%%%%%%%%%%%%%%%%%%%%%%%%%%%%%%%%%%%%%%%%%%%%
\begin{figure}[ht]
\begin{center}
\includegraphics[scale=0.6]{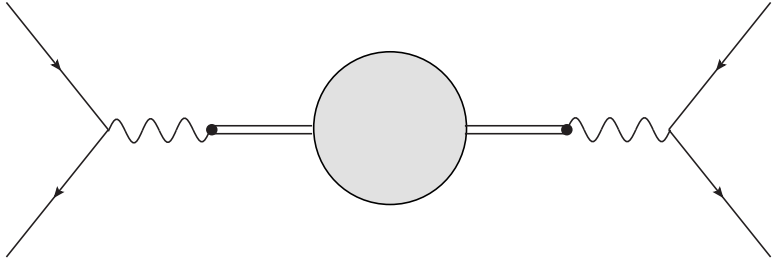}
\caption{A diagram to represent the matrix element  $\mathcal{M}(e^+e^-\rightarrow e^+e^-)$ for which one applies the optical theorem to obtain $\sigma(e^+e^-\rightarrow hidden)$.  The double lines denote  an intermediate vector resonance or resonances, depending on the model.  The blob denotes the current correlator $\tilde{C}'_1(s)$.  The cut is applied across the correlator.}
\label{eplusdiagram}
\end{center}
\end{figure}
%%%%%%%%%%%%%%%%%%%%%%%%%%%%%%%%%%%%%%%%%%%%%%%

In general gauge mediation, observable cross sections, for energies large enough to create on shell hidden sector states, may be extracted from the current correlators by application of the optical theorem:
\be
\sigma_a(\text{vis}\rightarrow \text{hidden}) =\frac{(4\pi \alpha)^2}{s}\frac{1}{2}\ \text{Disc} \ \tilde{C}_{a}(s).\label{scatterings1}
\ee
where $s$ is the Mandelstam variable, center of mass energy squared associated with the discontinuity across the cut. 
Conversely, in principle one could completely determine the current correlators from the experimental cross sections:
\be
i(16\pi^2 \alpha)^2\left[ \tilde{C}_{a}(s)- \tilde{C}_{a}(0)\right]=\frac{s}{\pi}\int_{s_0}^{\infty} ds'  \frac{\sigma_{a}(s')}{s'-s},
\ee
where $s_0$ is the end point of the cut and a sum over multiple cross sections is implicit.  Hence, there is an important relationship between sfermion masses and in-principle observable cross sections.    If one were to assume, as in QCD, intermediate resonances $\sigma(vis\rightarrow V_n\rightarrow hidden)$, then this vector meson saturated form factor may be pulled outside of the current matrix element and hence outside of the current correlator.
In other words, and to make clear the correspondence with generalised vector meson dominance and current matrix elements, we should associate
\be
\tilde{C}_1(s)\Pi_{\mu\nu}\sim \braket{h|J^{em}_{\mu}|0}\braket{h|J^{em}_{\nu}|0}^*
\ee
where $h$ labels the hidden sector fields and the matrix element couples to the external electromagnetic field.  The defining relationship between matrix elements is given by
\be
\braket{B|J_{\mu}^{em}|A}=\int d \sigma^2 \frac{\rho(\sigma^2)}{s-\sigma^2+ i\sigma\Gamma}\braket{B|J_{\mu}^{\rho_n}|A}=F(s)\braket{B|J_{\mu}^{\rho_n}|A}.\label{defining}
\ee
where we label $F(s)$ the form factor to simplify notation.  For scattering, the spectral function to use is $\rho(s)=m_n^2 .... $, for example as in \refe{specfunction}.

As a result, for a general resonance model, the contribution from intermediate s-channel resonances may be written
\be
\sigma_a(\text{vis}\rightarrow \text{hidden}) =\frac{(4\pi \alpha)^2}{2s}\int d \sigma^2 \frac{\rho(\sigma^2)}{s-\sigma^2+ i\sigma\Gamma}\int d \sigma^{'2} \frac{\rho^{}(\sigma^{'2})}{s-\sigma^{'2}-i\sigma \Gamma}\ \text{Disc}\  \tilde{C}'_{a}(s).\label{scatterings2}
\ee
With $\Gamma$ the decay width. The $\tilde{C}'_{a}(s)$ with a  $'$,  now being the correlator of current matrix elements relating vector mesons to hidden sector fields: 
\be
\tilde{C}'_1(s)\Pi_{\mu\nu}\sim \braket{h|J^{\rho_n}_{\mu}|0}\braket{h|J^{\rho_m}_{\nu}|0}^*.
\ee
This is pictured in figure \ref{eplusdiagram}.

Finally, to see that this is the correct result we display the canonical spectral function \refe{canonical}
\be
\rho(\sigma)=\sum_v \frac{m_v^2}{g_v}\delta(\sigma^2-m_v^2),
\ee
completely reproducing known results for generalised vector meson dominance. In this way one may obtain a cross section for these resonance models from the cross sections of \cite{Fortin:2011ad}.  The above result is obtained in an entirely \emph{four} dimensional construction by assuming a generalised form of vector meson dominance.  However,  a holographic interpretation can be made in which the form factors are associated with a bulk to boundary five dimensional propagator.  Finally applying this understanding of cross sections in reverse, one can then infer the form of the sfermion mass formula to be \refe{sfermionmasses}. 

To clarify this little further, one interprets the correlators of \cite{Fortin:2011ad}  to be associated with vector meson-hidden interactions and assume that the hidden sector has some  weakly coupled description such as minimal  gauge mediation with superpotential $W=X\Phi \tilde{\Phi}$. To compute the correlator ones starts with the cross section associated with the scalar current 
\be
J^a= \phi^{\dagger}T^a\phi-\tilde{\phi}T^a\tilde{\phi}^{\dagger}= \phi^{\dagger}_{+} T^a\phi_{-}+\phi^{\dagger}_{-}T^a\phi_{+}.
\ee
This will produce two hidden sector scalars of masses $m_+$ and $m_-$, 
\be
\sigma_0 (s) =(4\pi\alpha)^2 |F(s)|^2\frac{\lambda^{1/2}(s,m_+,m_-)}{8\pi s^2} |\frac{\mathcal{M}}{4\pi\alpha}|^2
\ee
 $\mathcal{M}$ is the corresponding matrix element for this process\footnote{Note that we have defined dimensionless spectral functions which account for a factor of $s$.}. The phase space or `triangle factor' is given by
\be
\lambda^{1/2}(s,m_1,m_2)=2\sqrt{s}|p^{ \! \! \! \to}|=\sqrt{[s-(m_1+m_2)^2][s-(m_1-m_2)^2]}\theta(s-(m_1+m_2)^2)
\ee
$\theta$ is the unit step function, for positive values of its argument.  One can use the optical theorem to obtain\footnote{see page 310  of  \cite{Itzykson:1980rh} }
\be
\frac{1}{2}\text{Disc}  \ \tilde{C}'_{0}(s)=\frac{1}{2} \left(\tilde{C}'_0(s+i\epsilon)-\tilde{C}'_0(s-i\epsilon)\right)=\frac{1}{4\pi s}\sqrt{s^2-4|M|^2+4|F|^2}.
\ee
The final cross section is given by 
\be
\sigma_0(\text{vis}\rightarrow \text{hidden},s) =(4\pi \alpha)^2|F(s)|^2\frac{1}{4\pi s^{2}}\sqrt{s^2-4|M|^2+4|F|^2}.
\ee
There may be additional kinematic factors for the other cross sections associated with $\tilde C'_a (s)$ for  $a=0,1/2,1$.  Also, in the two site model for instance, these results are modified to take account of $g_0$, $g_1$ gauge couplings at each lattice site.  It is straightforward to extend the results of \cite{Fortin:2011ad} in this way.   It is notable that for the resonance masses $m_n^2(1+2n)$, the cross sections can acquire an approximately scale invariant form \cite{Bramon:1972vv,Sakurai:1973rh}.   

Let us now emphasise the close relationship between these cross sections and  the spinless external states of hadrons in QCD, for definiteness we focus on $\sigma(e^+,e^-\rightarrow \pi_+,\pi_-)$.  The pions of QCD are in the adjoint of $SU(2)$ where the current is  $J^{\mu}_{\pi}=\overrightarrow{\pi}\times \partial^{\mu}\overrightarrow{\pi}$ whereas in this case the messengers are taken to be in the fundamental in this example.  This is meant as an instructive analogy.   To do this take $F\rightarrow 0$ and identify $M_{SUSY}$ with $M_{\pi}$.  In this case the phase space factor reduces to the simple form 
\be
\lambda^{1/2}(s,m,m)=[s(s-4m^2)]^{1/2}=s\sqrt{1-\frac{4m^2}{s}}\theta(s-4m_{\pi}^2).
\ee
The actual correlator of the pion current is 
\be
\tilde{C}_{\pi}(s) =\frac{2}{3p^2} \int \frac{d^4q}{(2\pi)^4}\frac{(p+k)_{\mu}k^{\mu}}{(p+k)^2+m_{\pi}^2}\frac{1}{k^2+m_{\pi}^2}
\ee
giving
\be
\text{Disc}\ \tilde{C}_{\pi}(p^2)=\frac{1}{12s^2}(s-4m_{\pi}^2)\lambda^{1/2}(s,m_{\pi},m_{\pi})
\ee
and therefore 
\be
\sigma(e^+,e^-\rightarrow \pi^+,\pi^-,s)=\frac{(4\pi\alpha)^2}{4\pi 12 s  }\left(1-\frac{4m^2}{s}\right)^{3/2}|F(s)|^2\theta(s-4m^2).
\ee
   One can straightforwardly extend this approach to many other cross sections, including higher spin states, and for instance if one assumes supersymmetry breaking sector baryons $B_h$,  then using \refe{defining} the typical scattering of off-shell standard model visible gauge fields $\gamma$ or $\lambda$ to $B_h$ will essentially take the form 
\be
\sigma^T_{\gamma^* B_h}(s)=(4\pi\alpha)|F(s)|^2 \sigma^T_{\rho_n B_h }(s)
\ee
where typically a multiplying factor $s/m_n^2$ is necessary for off-shell longitudinal gauge fields.

That this construction reproduces already known results for vector meson dominance cross sections is a powerful confirmation of this  approach. Furthermore, whilst the results may be motivated by ``5d" bulk to boundary propagators, the result may be thought of as entirely four dimensional.   This is simply a direct application of the ``modified current operator". It is also easy to calculate and simple to fit to (future) data.   Whilst these cross sections are associated with some scale $M$, typically hundreds of TeV and therefore out of reach of present accelerators, it is hoped these results give some important insights for model building of supersymmetry breaking. 

In \cite{Fortin:2011ad}   the operator product expansion  (OPE) was used to determine the sfermion masses of \cite{Martin:1996zb}.  Essentially the OPE is a $1/s=M^2/p^2$ expansion.  For models where resonances contribute to the mediation, the dominant contribution to the integral for sfermion masses is for small $s$, where an $p^2/M^2$ expansion is more appropriate.    The OPE expansion still correctly determines the sfermion soft masses in the limit $m_{\rho}\gg M$, although unfortunately this is not the region of interest.

\section{The A-term}
In this section we wish to show that the feature of the spectral function also arises for A-terms.  The leading gauge mediated A-term contribution is the third generation $a_t Q\bar{u}H_u$.  In resonance mediation, with no additional features, it is given by 
\be
a_t=- \frac{8y_t}{3} (4\pi\alpha_3)^2\int \frac{d^4p}{(2\pi)^4}\int d^2 \sigma \frac{\rho(\sigma^2)}{s-\sigma^2}\int d^2 \sigma' \frac{\rho(\sigma^{'2})}{s-\sigma^{'2}}\frac{MB_{1/2}(p^2/M^2)}{(p^2-m_t^2)^2}
\ee
As a two loop diagram it is severely suppressed if the sfermions mass formula is suppressed, at the high scale $M$, by the same form factor as the sfermion masses.   For the case of a simple two site quiver and minimal gauge mediation sector, it may be evaluated exactly using identities from \cite{vanderBij:1983bw,Ghinculov:1994sd}. 

 In the 4d limit in which the resonance scale is much heavier than $M$, and with the minimal model $W=X\Phi\tilde{\Phi}$ behaves as
\be
A_t(4d)=-y_t \frac{8}{3} (\frac{\alpha_3}{4\pi})^2  m_0 \left(h (a,b)-h(a,c)\right)
\ee
where $h(a,b)$ is a function given in \cite{vanderBij:1983bw,Ghinculov:1994sd}.
The function $h$ is given by the integral
\be
h(a,b)= \int_0^1 dx  \left( 1+ {\rm Li}_2 (1-\mu^2) -\frac{\mu^2}{1-\mu^2}  \log \mu^2 \right) \, 
\ee
The dilogarithm is defined as ${\rm Li}_2(x)=-\int_0^1 \frac{dt}{t}\log(1-xt)$  with
\be
\mu^2=\frac{a x + b(1-x)}{x(1-x)} \, \ \ , \ \   a=m_1^2/m_0^2   \ \ , \ \ b=m_2^2/m_0^2.
\ee
and 
\be
a=\frac{m^2_0}{m^2_t} \ \ ,  \ \ \ b=\frac{m^2_+}{m^2_t}  \ \ \ \,  \ \ \ c=\frac{m^2_-}{m^2_t}
\ee
Taking the leading log pieces  with $y=M/m_t$ and $x=F/M^2$
\be
(h (a,b)-h(a,c))=\left(\log[1-x]-\log[1+x] \right)\log[y]+...
\ee
In the limit of massless $m_t$  it behaves as
\be
\left(h (a,b)-h(a,c)\right)\sim -x,
\ee
which vanishes as $F\rightarrow 0$. The $A_{t}$ is a two loop effect, which is further suppressed by the presence of resonances.  A viable model may therefore require further extensions to overcome this issue, many have already been proposed which involve direct couplings between Higgs multiplets and the supersymmetry breaking \cite{Giudice:1998bp,Craig:2012xp} or by modifying D terms using vector mesons \cite{Auzzi:2011eu}.   This result is of course well known but it is  included it to highlight the appearance of the form factors, which are in principle observable from cross sections, in various loops.

\section{Holography and The Generating Functional}\label{holo}
In this section we  wish to apply a holographic model as an effective theory of a strongly coupled large $N_c$ SQCD sector that breaks supersymmetry.  We assume that this $SU(N_c)$ has in addition some global symmetry $SU(N_F)$ which is weakly gauged and associate with the standard model gauge groups. This demonstrates much of the utility of general gauge mediation \cite{Meade:2008wd} and is a concrete model from which to extract the various form factors for both sfermion masses and cross sections.   This section overlaps with the hidden local symmetry program of AdS/QCD  \cite{Rho:2008zz} and shares features with \cite{Nomura:2004zs,Abel:2010uw,McGarrie:2010yk,Abel:2010vb}, which we will abbreviate to AdS/\cancel{SUSY}.

In analogy to hidden local symmetry models \cite{Rho:2008zz,Abel:2012un}, we apply a type of Wilsonian matching, where we wish to analyse a system which is described by an effective theory
\be
Z_{SSM} [\mathbb{J}]\times Z_{SQCD} [\mathcal{J}]|_{E\ll \Lambda}\rightarrow Z_{SSM} [\mathbb{J}]\times Z_{AdS}[\mathcal{O}] \times Z_{IR}[J].\label{correspondence}
\ee
The effective action is the bare action or the classical action at the scale $\Lambda$.  The theory on the left hand side is four dimensional.  The effective description on the right hand side is five dimensional in which we have introduced an additional dimension $z$.  We make the assumption that we can match the observable physics on both sides  for energies around the matching scale $\Lambda$.  Then at energies $E\ll \Lambda$ we wish to use the generating functional on the right hand side to compute physical quantities.  In particular this means that instead of  attempting to solve certain equations of motion with current sources on the IR brane, we can use perturbation theory.

The generating functional for the supersymmetric standard model we will label by
\be
Z_{SSM}[\mathbb{J}]=\int D[SSM, V(x)=\tilde{V}(x,z=L_0)]e^{iS(SSM, V(x);\mathbb{J})}
\ee
Where we take the $SSM=QUDLEH_u H_d$, the supersymmetric standard model fields,  which we locate on the UV boundary (outside the AdS system).

The large $N_c$ strongly coupled supersymmetry breaking sector we will label SQCD:
\be
Z_{SQCD}=\int D[Q,\tilde{Q},G]e^{iS(Q,\tilde{Q},G ; \mathcal{J})}
\ee
We assume that at energies $E\ll\Lambda$ it admits a dual ``slice of AdS''
\be
Z_{AdS}[\mathcal{O}]=\int D[\tilde{V},\Phi_{Adj}]e^{iS(\tilde{V}, \Phi_{Adj};\mathcal{O})}
\ee
The $\tilde{V}+\Phi_{Adj}$ are the superfields of $\mathcal{N}=1$ SYM in 5d \cite{Hebecker:2001ke}.  We take the supergravity background to be static but those degrees of freedom should really be included in an analysis of $hidden\rightarrow hidden$ scattering, for example.    For our purposes we are interested in modeling the behaviour of the gauge field in this background.  Importantly, we have chosen the $\tilde{V}$ to have positive parity (and therefore a massless zero mode spectrum) and $\Phi$ to have negative parity and no massless modes.  This choice is consistent with phenomenology.  The supersymmetric action plus parity completely determines the field profiles and the relative field profiles between each field in the same multiplet. 

In the dual description we expect that the actual supersymmetry breaking is located on the IR brane
\be
Z_{IR}[J]=\int D[X, \Phi, \tilde{\Phi}, V_{IR}=\tilde{V}(x,z=L_1)]e^{iS(\Phi, \tilde{\Phi}, V_{IR};J)}.
\ee
The field $X$ is the supersymmetry breaking spurion, $\Phi,\tilde{\Phi}$ are the messenger fields, that are coupled directly to the spurion $X$ and from which the global currents of $J$ are extracted.  In principle one may locate the supersymmetry breaking fields in the bulk and remove the IR brane.  We choose the above construction for visual clarity.  The hidden sector is therefore the combination of $Z_{AdS}[\mathcal{O}]\times Z_{IR}[J] $ and in particular the resonances of $\tilde{V}+\Phi_{Adj}$ are composites of the hidden sector with standard model quantum numbers.

In this model both the UV and IR brane fields are dynamical and must be integrated over.  At each point along  the AdS direction is the  field $\tilde{V}(x,z)$  and specifically its UV limit $V$ is the external gauge field that couples to the supersymmetric standard model fields.  At the IR brane, the supersymmetry breaking effects are encoded in $J$ and couple to $V(z=L_1)$ \cite{ArkaniHamed:2000ds, Randall:2001gb,Agashe:2002jx,Chacko:2003tf}.

 In terms of the $AdS_5$ direction $z$, this corresponds to $z=L_1$.  In the first instance the supersymmetry breaking effects, which are encoded in the currents labelled $J$, do not couple directly to $A_{\mu}(x,z=L_0)$ but instead to $A_{\mu}(x,z=L_1)$, a characteristic of locality in theory space.   Furthermore, as $\alpha_{SM}\rightarrow 0$ all effects, such as soft masses, of supersymmetry being broken must vanish. 

In particular we are not breaking supersymmetry using boundary conditions.  Instead we have a dynamical  sector located on the IR brane which spontaneously breaks supersymmetry and for which we can clearly define a goldstino field.  In addition this will allow us to compute cross sections to messenger fields.   We also make the requirement that our messenger sector satisfies the sum rule $\text{Str}\mathcal{M}^2=0$ before gravity effects are taken into account.

\subsection{Holographic Gauge Fields}
In this section we will define the relevant parts of the gauge theory living inside the $Z_{AdS}[\mathcal{O}]$.  We will take $\mathcal{N}=1$ 5D Super Yang Mills on a slice of $AdS_5$ background \cite{Hebecker:2001ke,Abel:2010vb,Bagger:2011na} with coupling $g_5$.  We will then couple the supersymmetry breaking sector located at $Z_{IR}[J]$.

Starting with the conformally flat metric  with mostly minus signature
\be
ds^2= a^2(z)( \eta^{\mu\nu}dx_{\mu}dx_{\nu}-dz^2)
\ee
where
\be
a= \left(\frac{R}{z}\right)=\left(\frac{1}{\omega z} \right)    
\ee
$R\equiv L$ is the AdS curvature radius, with $L_0$ the position of the UV brane and $L_1$ the position of the IR brane.  We may relate $R^4=4\pi l_{s}^4g_s N_c$, where the string coupling is related $g_s=g^2_{YM}$ to the boundary $SU(N_c)$ gauge coupling $g_{YM}$. $l^2_s=\alpha'$ is the string length.  The 't Hooft coupling is $\lambda_t=g^2_{YM} N_c$.  See also \cite{McGarrie:2012fi,McGarrie:2013hca}.

 One may interchange between conformal coordinates and warped coordinates of ``proper distance'' using 
\be\frac{z}{R}=e^{ky}     \ \ \ 
\partial_y= a^{-1} \partial_z    \ \ \ \ dy= a(z) dz  \ \ \ \sigma = \ln \left(\frac{z}{R}\right)
\ee
where $k$ is the AdS warp factor. The kinetic terms for the super Yang Mills action is given by
\be
\int d^5x \ \text{tr}\left[- \frac{1}{2}F_{\mu\nu}F^{\mu\nu}(z)+i\left(\frac{R}{z}\right)^3 \bar{\lambda}\sigma^{\mu}\partial_{\mu}\lambda(z) +\frac{1}{2} \left(\frac{R}{z}\right)^4 D^2(z)\right].\label{warpedaction}
\ee
The bulk vector super fields have a Kaluza-Klein decomposition \cite{ArkaniHamed:2000ds,Gherghetta:2000qt}
\be
V(x,z)=\sum_n V_n(x) f_n^{(2)}(z) \ \  \text{and} \ \  \Phi(x,z)=\sum_n \Phi_n(x) g_n^{(4)}(z).
\ee
The vector superfield and chiral superfield having even and odd parity, respectively.  Once an IR and UV brane is implemented the bulk fields will obtain a KK spectrum.  As a result the parities of the bulk fields have been chosen so that massless modes such as $\chi\subset \Phi$ do not make the spectrum of the model phenomenologically incompatible.  If this field were given even parity and a bulk $\mathcal{N}=2$ SYM with 8 supercharges, then a soft Dirac mass could in principle be generated between $\chi$ and $\lambda$, but one may require exact $\braket{\Phi}^a_0 T^a\equiv 0$. We have chosen $\Phi$ to have odd parity and so no Dirac mass may arise.

The solutions of the KK wave functions are given by
\be
f_n^{s}(z)=\frac{1}{N_n} a^{(s/2)} [  J_{1}(m_n z)+ b_{\alpha}(m_n) Y_{1} (m_n z)]
\ee

\be
g_n^{s}(z)=\frac{\epsilon(z)}{N_n} a^{(s/2)} [  J_{0}(m_n z)+ b_{\alpha}(m_n) Y_{0} (m_n z)]
\ee
where $\epsilon(z)$ is a sign function. The positive parity boundary conditions on the branes fix
\be
b(m_n)= - \frac{J_1(m_n R)}{Y_1(m_n R)}.
\ee
The orthonormality condition is given by 
\be
\int  a^{(s-2)}  f^{s}(z)f^{s} (z)dz=\delta_{nm}
\ee
It is often useful to use 
\be
\partial_z g^{(4)} \partial_z g^{(4)} \simeq m^2_n f^{(2)} f^{(2)}.
\ee
One can then define a bulk AdS two point function for the vector fields,
\be
G(p^2;z,z')= \sum_n\frac{f^{(2)}_n(z) f^{(2)}_n(z')}{p^2-m_n^2}  \label{warpedformfactor}
\ee

\subsection{Soft Masses and Cross Sections}
Suppose one now integrates out $Z_{IR}[J]$ (one may later integrate out $Z_{AdS}[\mathcal{O}]$).  Introducing a specially normalised source term 
\be
g_5\int d^5x \delta(z-L_1)\left( a^2\hat{J}D - a^{3/2}\hat{j}^{\alpha}\lambda_{\alpha}-a^{3/2}\hat{\bar{j}}_{\dot{\alpha}}\bar{\lambda}^{\dot{\alpha}} -\hat{j}^{\mu}A_{\mu}  \right)
\ee
where $a=R/z$. The presence of currents will change the equations of motion to include source terms. Integrating out this sector, this will  generate an effective action for the gauge supermultiplet that couples to $V(x,z=L_1)$:
\be
\mathcal{L}_{eff}=\frac{g^2_{5d}}{2}\tilde{C}_0(0)D^2(z)\left(\frac{R}{z}\right)^4-g^2_{5d} \tilde{C}_{1/2}(0)i\lambda\sigma^{\mu}\partial_{\mu}\bar{\lambda}(z)\left(\frac{R}{z}\right)^3-\frac{g^2_{5d}}{4}\tilde{C}_{1}(0)F_{\mu\nu}F^{\mu\nu}(z)\label{warpedeffective}
\nonumber \ee
\be
-\frac{1}{2}g^2_{5d} M\tilde{B}_{1/2}(0)(\lambda \lambda(z) +\bar{\lambda}\bar{\lambda}(z) )\left(\frac{R}{z}\right)^3+ ...\label{effectiveaction}
\ee
the ellipses denote higher order terms in $g^2_{5d}$ and higher orders in momentum, and we take $z=L_1$.   The first line of \refe{effectiveaction} corresponds to wavefunction renormalisation caused by integrating out the fields $X,\Phi,\tilde{\Phi}$, and will in turn cause a change in the beta function $\Delta b_{IR}$ but this will not have much effect above the scale $E\sim1/L_1$.  Canonically normalised currents have been used such that the mass scales are given by
\be
\left(\frac{R}{z}\right)^2F=\hat{F} \ \ \  \left(\frac{R}{z}\right)M=\hat{M}.
\ee
In this way one may compute the loops of messenger fields.  The masses of the messenger fields are now 
\be
m^2_{\pm}=\hat{M}^2\pm \hat{F}  \ \  \text{and}   \ \   m_0^2=\hat{M}^2.
\ee
One then uses this effective action to generate soft masses for the SSM located at $z=L_0$.

\subsubsection{Soft Masses in General}
The gaugino soft mass is given, after canonically normalising the kinetic term of $\lambda$ in \refe{warpedaction} relative to the soft mass of \refe{warpedeffective}, by
\be
m_{\lambda n,m}=(4\pi\alpha)\hat{M}\tilde{B}_{1/2}(0) f_n f_m+c.c.
\ee
as it is at $p_{ext}=0$ it is unnaffected by form factors: the physical gaugino mass is the shift pole in the propagator,  whilst there are form factors in cross sections to $\sigma'_{1/2}\sim G(z,z') G(z,z') \ \text{Disc}  \hat{M}\tilde{B}'_{1/2}$, but the masslesss mode of these form factors are simply part of the geometric sum of mass insertions that make up this shifted pole and therefore do not effect the identification of the soft mass.  There are also \emph{supersymmetric} Dirac masses, associated with the Kaluza Klein modes of the gauginos between the fermions in $\Phi$ and those of $V$ i.e. $\partial_z\lambda\partial_z\chi$. In particular the field $\Phi$ is chosen to have negative parity on at least the UV or IR brane so that a massless fermion $\chi_0$ does not appear in the spectrum.

The sfermion masses are given by 
\be
m_{\tilde{f}}^2=- \left(4\pi\alpha_{5d}\right)^2\! \! \! \int \! \! \frac{d^4 p}{(2\pi)^4} G(p^2;L_0,z) G(p^2;z,L_0) p^2 \Omega \left(\frac{p^2}{M^2}, \frac{R}{z}\right).
\ee
where we take $z=L_1$.  It is clear that the sfermion masses will be suppressed from the intermediate resonances. 
\subsubsection{Soft Masses with a Minimal Messenger Sector}
 If one were to take a minimal messenger model $W=X\Phi\tilde{\Phi}$ to evaluate the current correlators then one obtains in the limit $m_{n}\ll M$
\be
\Omega\left(\frac{p^2}{M^2}, \frac{R}{z}\right)=-\frac{1}{(4\pi)^2}  \frac{2}{3}\left|\frac{F}{M^2}\right|^2 h(x) 
\ee
\cite{McGarrie:2010kh} which is independent of $p^2$ ($h(x)\sim 1$).  In this case the sfermion masses are given by 
\be
m^2_{\tilde{f}}\sim\frac{2}{3} \left(\frac{\alpha_{5d}}{4\pi}\right)^2  \left(\frac{R}{z}\right)^2 \left| \frac{F}{M}\right|^2\left|\frac{1}{\hat{M}}\right|^2\times \int dp \  p^5 G(p^2;L_0,z)G(p^2;z,L_0)
\ee
Written in this way, one can see the contribution directly from dimensional analysis and the part associated with the bulk propagator/form factor associated with the resonances. In principle the supersymmetry breaking is located at $z=L_1$ however we have left it as a free variable in the above expression. The integral will supply a scale $m_{kk}$ after a change in varables in $p$. It is not completely free, however as the correlators have been expanded in a particular limit. Following the intuition of Hybrid mediation \cite{Auzzi:2010mb}, one could change the scaling behaviour by adjusting $m_n, M$ and also the location along $z$ of the supersymmetry breaking closer to the UV boundary and obtain an expression of the form
\be
m^2_{\tilde{f}}\sim \frac{2}{3}\left(\frac{\alpha_{5d}}{4\pi}\right)^2\left(\frac{R}{z}\right)^{\tau}  \left| \frac{F}{M}\right|^2\left|\frac{1}{\hat{M}}\right|^{\tau} \times \int dp \  p^5G(p^2;L_0,z)G(p^2;z,L_0)
\ee
where $\tau \in [0,2]$. $\tau$ is actually a function of all the scales and there is an important leading coefficient,  but none the less it seems promising that one could obtain a Hybrid regime where effectively only the first few resonances contribute\footnote{A two site quiver is always ``flat''.   In Deconstructed AdS models \cite{deBlas:2006fz,Falkowski:2006uy} one requires at least 3 sites and therefore two resonances of $A_{\mu,i}$ $i=1,2,3$.  This may suffice as a reasonable model of the full setup. } and $\tau\sim 1$.

In the limit in which $m_n\gg M$  only the massless modes mediate supersymmetry breaking and the soft masses are given by 
\be
m^2_{\tilde{f}}= 2 \left(\frac{\alpha_{5d}}{4\pi}\right)^2 \left| \frac{F}{M}\right|^2 \left(\frac{R}{z}\right)^{2}
\ee
where again $z$ is set to the position of the effect action $z=L_1$.  In all cases the gaugino  soft mass terms are given by 
\be
m_{\lambda_{n,m}}=2 \frac{\alpha_{5d}}{4\pi}\frac{F}{M}\left(\frac{R}{z}\right)f_{n}f_m g(x)
\ee
where for $x=F/M^2$  and for $x<1$, $g(x)\sim 1$.  

In summary, within this type of construction soft masses are completely calculable.  For heavy resonances $m_n\gg M$ the model encodes a natural hierarchy of scales relative the $M_{Pl}$, however the soft masses and scattering cross sections will appear four dimensional and the effect of resonances will be limited.  Conversely in the limits $m_n\sim M$ and  $m_n\ll M$ sfermion mases will be suppressed and gaugino masses unchanged.

\subsubsection{Cross Sections}
Additionally it is interesting to see the effect upon scattering cross sections \cite{Fortin:2011ad}.  The form factor is associated with the bulk to boundary propagator \cite{Erlich:2005qh}.   Taking a minimal model of the supersymmetry breaking sector we look at the \newline $e^{+}e^{-}\rightarrow \{\gamma_n\}\rightarrow \text{hidden sector matter}$, where  one must consider all ($n$) photon resonances as intermediate states. This corresponds to computing $\tilde{C}_1(s)$.  The corresponding matrix element is given by\footnote{To be compared with page 616 Chaper 18 of \cite{Peskin:1995ev}}
\be
i\mathcal{M}(e^+e- \rightarrow e^+e^-)=(-ie)^2 \bar{u}(k)\gamma_{\mu}v(k_+) \frac{-i}{s}F(s)(i\tilde{C}'_1(s))\frac{-i}{s}F^*(s)\bar{v}(k_+) \gamma_{\nu}u(k)
\ee
with $F(s)$ being the corresponding form factor.  One then applies the optical theorem which acts on $\tilde{C}'_1(s)$.  It turns out that form factors in AdS are more naturally formulated in terms of the holographic basis  instead of the Kaluza Klein basis,  the specific details may be found in \cite{McGarrie:2012fi}  but essentially the form factor is given by an amputated Green's function or equivalently a bulk to boundary propagator. The form factor of this model is given by taking the bulk to boundary propagator 
\be
K(p^2, z')=g_5\sum_n \frac{F_n \psi_n(z) }{p^2-m_n^2}
\ee
This encodes a sum of monopole contributions of an infinite tower of vector mesons with decay constants  $F_n= \braket{0|\mathcal{O}|\rho_n }$ for the nth meson of mass $m_n$.  Defining 
\be
F(p^2)=\int d z a(z) K(p,z) \phi(z)\tilde{\phi}(z)
\ee
where $ \phi(z),\tilde{\phi}(z)$ are the wavefunctions of the messenger fields (which in this case are trivial and include a $\delta(z-L_1)$.  In general one may define
\be
g_{\rho \phi\tilde{\phi}}=\int dz a(z) \psi(z) \phi(z)\tilde{\phi}(z)
\ee
in analogy to pion scattering.  The size of the messenger field's charge distribution is defined from this form factor:
\be
\braket{r^2}=6\frac{\partial}{\partial s} F(s)_{s=0}=-\sum_n \frac{6 g_5 F_n \psi(L_1)}{m_n^4}.
\ee
It may be evaluated by taking $F_n=F$, $\psi_n(L_1)=d(-1)^n$  and $m_n^4 =n^4 m_{\rho}^4$
to obtain
\be
\braket{r^2}=\frac{42 g_5F d  \pi^4}{720m_{\rho}^4}
\ee
as $F_n\sim m_n^2/g_5 $,  $\braket{r^2}$  is $O(1/m^2_{\rho})$ and can be compared with \refe{r}.  The two site quiver model of section \ref{A Toy Model} has some universal couplings associated with a hidden local symmetry as observed in \cite{Komargodski:2010mc}.  These slice of AdS models do not reproduce the full set of relations between $g_{\rho \phi\tilde{\phi}}$ and $g_{\rho \rho \rho }$ however they do relate separately for $g_{V\phi\tilde{\phi}}$ and for  $g_{VVV}$.

Following \cite{Fortin:2011ad} one may take a minimal messenger sector where the cross section is given by 
\be
\sigma_{1}(e^+,e^-\rightarrow \text{hid})=|F(s)|^2\frac{(4\pi\alpha_{5d})^2}{48\pi s^2 } 
\ee
\be
\times [(s-4m^2_+)^2\lambda^{1/2}(s,m_+,m_+)+ (m_+\rightarrow m_-)+4(s+2m_0)\lambda^{1/2}(s,m_0,m_0)]\nonumber
\ee
where this result is related to the $\tilde{C}'_1(s)$ correlator at $z'=L_1$.   In this way the effect of intermediate resonances are quantified and additionally the cross sections obtain a scaling behaviour.   Similar examples can be found for holographic models of pion scattering \cite{DaRold:2005ju}.

\subsubsection{Messengers In The Bulk}
One may imagine a situation where, for definiteness, the messenger fields $\Phi,\tilde{\Phi}$ live in the bulk and are associated with positive parity component of bulk hypermultiplets.  Using the \emph{loose} analogy of AdS/QCD models, where the pions are modelled by a bulk adjoint field $X(x_{\mu},z)$ \cite{Erlich:2005qh}.    Due to the warping we expect these fields to be mostly localised towards the IR, quantified by the value $c$ of their five dimensional wavefunction $\Phi_n(z,c)$ \cite{Gherghetta:2000qt} and on the four dimensional side of the correspondence \refe{correspondence}, there will be operators which correspond to these bulk fields.  The supersymmetry breaking currents $\mathcal{J}(z,x)$ are extracted from their kinetic terms. 

In this case we must integrate over $z$, so we may pull this factor out by dimensional analysis
\be
\Omega\left(\frac{p^2}{M^2}, \frac{R}{z}\right)= g(z)\Omega \left(\frac{p^2}{M^2}\right)
\ee
As the currents are loops of messenger fields they will now be position dependent and also preserve incoming and outgoing $p_5$ momenta.   The sfermion mass formula is then given by
\be
m_{\phi}^2=- \sum_n (4\pi\alpha_{5d})^2\! \!  \int \! \frac{d^4 p}{(2\pi)^4} \frac{f^{(2)}_n(0) f^{(2)}_n(0)p^2}{(p^2+m_n^2)^2}   \  \Omega\left(\frac{p^2}{M^2}\right)  \int f^{(2)}_n(z)f^{(2)}_n(z) g(z)dz
\ee
It should be clear that even without specifying the exact form of the current correlators the leading sfermion mass contribution will be suppressed.  The relevant cross sections of section \ref{crosssections} will involve a single summation over resonances instead of a double sum.  This single sum instead of double sum would appear in cross sections too, as an interference effect, essentially pulling out the sum on $n$ in the form factor $\sum_n|F_n(q)|^2$.

 Finally, if we located the correlators in the UV $\tilde{C}_{a}(z=L_0)$ then only the massless modes, whose wavefunctions are flat, would mediate the supersymmetry breaking at all. This corresponds to a limit in which the external field no longer mixes with the resonances.

\subsection{Computing Correlators of $\mathcal{O}$}
There are many operators represented by $\mathcal{O}$.  These currents are associated with the broken CFT or AdS background  \cite{ArkaniHamed:2000ds} equivalent to the bulk gauge field self couplings \cite{Goldberger:2002cz,Randall:2001gc,Randall:2001gb}, where essentially resonances of $V_n$ and $\Phi_{Adj,n}$ run in the loop, and the coupling of the gauge field to other bulk fields. All these will generate contributions to the one loop vacuum polarisation amplitude of the external gauge field and therefore to the beta function, $\Delta b_{bulk}$,  or running gauge coupling of the external field.
% for example:
%\be
%\int d^4x e^{iq.x}\braket{\mathcal{O}_{\mu}(x)\mathcal{O}_{\nu}(0)}\sim %\sum\frac{[\partial_{z}\partial_{z'}G(q^2; z,z')]}{(q^2-m_{n}^2)m_{n^2}}\Pi_{\mu\nu}
%\ee
The effective action associated with these contributions will be analogous to \refe{effectiveaction} but located at the UV boundary.   The ``super-traced'' combination \refe{supertrace} of this current multiplet vanishes \cite{Meade:2008wd} however, as they contain no supersymmetry breaking effects\footnote{up to two loop order in the gauge coupling. There are three loop (and higher) contributions for instance as all the Gaugino (and kk mode) poles will be shift by a soft mass\cite{Lee:2010kb}.} within this type of supersymmetry breaking.  
The supersymmetry breaking affects are associated with the $J$ multiplet located at $z=L_1$, which is still a one loop correction to the external gauge field but which first descends down to couple to the relevant operators.  This is much the same as for lepton annihilation to pions in models of AdS/QCD.  The experimentally observed vacuum polarisation amplitude that couples to the external fields is to all orders in the hidden sector and that includes the contributions from currents contained in $\mathcal{O}$, $\mathbb{J}$ and $J$.  Such an effective action could be achieved by completely integrating out the IR and bulk generating functionals $Z_{AdS}[\mathcal{O}] \times Z_{IR}[J]$, leaving only 
\be
Z_{SSM}[\mathbb{J}]\times e^{i S_{eff}(V_{0})}.
\ee
%As a point in principle, one may decimate the generating functional by 
%\be
%Z_{BULK}=\prod_{i=1}^{\infty}Z[,z_i]
%\ee
%and integrate out the generating funcional for each $z_i$.

\subsection{A smoother transition?}
Finally let us discuss an application of Holography closer to the soft wall or unparticle scenario, where instead of an IR brane cutoff we wish for a smoother regulator associated with a $z$ dependent dilaton profile \cite{Rajaraman:2008bc}.  We can then imagine locating the supersymmetry breaking peaked at some definite value of $z=L_1$.  First setting the gauge couplings to zero, the propagators of the gauge fields are determined entirely by the geometry and their coupling to the dilaton.  Switching the gauge coupling back on, we can work perturbatively to generate suitable soft masses.  To obtain the results for a soft wall model from the above results, one may simply replace $G(z,z')$ with that obtained from the soft wall equations of motion.

\section{Discussion and Conclusion}
In this paper we have explored a natural extension of general gauge mediation to cases where the mediating gauge fields, instead of having a two point function of a free theory, have  a two point function described by a non trivial spectral function.  We have shown that this covers a very broad class of models, including those that involve a strongly coupled hidden sector, approximate conformality or Kaluza-Klein modes associated with an extra dimension.    Despite covering a broad class of models some general features may be extracted: 
 even within \emph{entirely four dimensional} models of a strongly coupled hidden sector, it is likely that one obtains suppressed scalar soft masses relative to gaugino masses at the high scale $M$ before RG running.  This is inferred by assuming that some hidden sector resonances will share the gauge field quantum numbers of the standard model and then vector meson dominance may ensue.  Of course these results are dependent on the assumption of a generalised form of vector meson dominance and we are mindful to point out that not all strongly coupled models exhibit this phenomena: some models may have straightforward magnetic duals.  

However within these models, typically one must compare the resonance scale with the scale $M$  of the hidden sector and for a sufficiently low resonance scale sfermion masses are suppressed at leading order and one observes a scaling in the  cross sections  of visible matter to hidden sector messenger fields.

It seems that to obtain quantitive predictions for effects on scattering cross sections and soft masses, particularly from string models, it may be useful to simply integrate out \cite{Hirn:2005nr,Harada:2010cn} all the higher states and extract the lowest resonance masses and relevant form factors.  An interesting consequence of this is that, if one starts from the holographic model above and integrates out all the higher resonances one may hope to arrive at a  quiver model very similar to that obtained starting from the magnetic description of certain Seiberg dual models explored in \cite{Green:2010ww,McGarrie:2010qr,McGarrie:2011dc,Auzzi:2010mb,Abel:2012un}.

We also identified for the simplest quiver the ``supercurrent field identity'' and ``modified current operator'' which are useful tools for interpreting resonances as they are part of the vector dominance model and showing how these are  related to the operator-field correspondence of the AdS/CFT perspective.   In a general sense these resonance mediation models capture the same features that were found with photon interactions to hadrons, so it is rather appealing and well motivated to build these features in to gauge mediation.  We believe that this is a powerful confirmation of this approach. 

One point we hope to have emphasised is that the general gauge mediation programme \cite{Meade:2008wd} is not simply a parameterisation of soft terms of the MSSM, it also encodes scattering cross sections \cite{Fortin:2011ad} of $visible\rightarrow hidden$.  In this paper we have extended these cross sections to general \emph{four dimensional} strongly coupled models and five dimensional models, in which there are intermediate resonances. In particular  the use of different types of integrals of the cross sections have been explored, to in principle reveal Schwinger terms as signs of vector meson dominance. 
It would be interesting to extend this work to $hidden \ sector \rightarrow hidden  \ sector$ scattering: whilst most of the examples in this paper have focused on AdS/CFT type constructions one can't help but feel that analysing the current correlators and scattering cross sections using Regge theory \cite{Collins:1977jy} may be as worthwhile.

\paragraph{Acknowledgements} 
MM is funded by the Alexander von Humboldt Foundation.  MM would like to thank Felix Brummer, Andreas Weiler, Mathias Garny, Daniel C. Thompson and Alberto Mariotti for fruitful discussions.  The diagrams are drawn in JaxoDraw  \cite{Binosi:2003yf,Binosi:2008ig}.

\appendix

\bibliographystyle{JHEP}
\bibliography{ADSOPE.bib}

\end{document}